\documentclass[twocolumn,showpacs,preprintnumbers,amsmath,amssymb,prb]{revtex4}

\usepackage{graphicx}
\usepackage{dcolumn}
\usepackage{bm}

\begin{document}

\preprint{}

\title{Discrete transverse superconducting modes in nano-cylinders}

\author{J. E. Han and Vincent H. Crespi}
\affiliation{Department of Physics and Materials Research Institute, 
The Pennsylvania State University, 
University Park, PA 16802-6300}

\date{\today}

\begin{abstract}
Spatial variation in the superconducting order parameter becomes
significant when the system is confined at dimensions well below the
typical superconducting coherence length.  Motivated by recent
experimental success in growing single-crystal metallic nanorods, we
study quantum confinement effects on superconductivity in a
cylindrical nanowire in the clean limit.  For large diameters, where
the transverse level spacing is smaller than superconducting order
parameter, the usual approximations of Ginzburg-Landau theory are
recovered. However, under external magnetic field the order parameter
develops a spatial variation much stronger than that predicted by
Ginzburg-Landau theory, and gapless superconductivity is obtained
above a certain field strength.  At small diameters, the discrete
nature of the transverse modes produces significant spatial variations
in the order parameter with increased average magnitude and multiple
shoulders in the magnetic response.
\end{abstract}


\maketitle

\section{Introduction}

Recent developments in nanofabrication techniques allow access to new
physical regimes where various intrinsic order parameters interact
with a tuneable confining environment. Such order parameters cover an
array of diverse physical systems, such as ferromagnets\cite{stoner},
quantum dots\cite{averin}, molecular electronics\cite{molelec},
photonic crystals\cite{sjohn},
superconductors\cite{graybeal,ralph,tinkham} etc.  Quantum-confined
superconductivity is particularly interesting for its macroscopic
quantum nature; its well-understood microscopic mechanism can also
serve as a platform for studies of other many-body nanoscopic quantum
confinement effects.

Since the advent of BCS theory, a great deal has been understood in both
the microscopic and phenomenological aspects of superconductivity for
conventional phonon-mediated pairing systems. Theories have been
immediately applied with great success to small
superconductors~\cite{tinkham,degennes} of various
geometry~\cite{suderow,misko}.  Particularly useful has been the
phenomenological Ginzburg-Landau theory, which describes
superconductivity directly in terms of the superconducting order
parameters without appealing to an underlying electronic basis.

Partly due to its enormous success, however, the Ginzburg-Landau
theory has sometimes been applied beyond its strict regime of
validity, especially in systems of small size.  Part of the
justification for this has been that experimental samples have often
been disordered or polycrystalline, in which case confinement effects
are less pronounced than they are for single crystals.
Well-established dirty-limit theories for small superconductors with
strong disorder~\cite{maki_parks} describe fascinating physics, such
as gapless superconductivity, down to nanometer scales.  In the clean
limit of microscopic BCS theory, where the mean free path $\ell$ is
longer than the coherence length $\xi_0$ ($\ell\gg\xi_0$), the
superconducting behavior of small samples is very different from that
in the dirty limit ($\ell\ll\xi_0$)\cite{strassler,misko}.  Recent
experimental techniques for producing high quality {\it single
crystalline} nanostructures through electrodeposition into extended
nanopores\cite{tian} demands a re-examination of the phenomenology of
superconductivity in such systems, working from microscopic theories.
We specifically aim to investigate the often-overlooked spatial
structures of the superconducting order parameter in the confined
direction by directly solving the Bogoliubov-de Gennes equation and
comparing with other theories.

For the last few decades, work on one-dimensional superconductivity
has mostly focused on fluctuation effects~\cite{giordano,tinkham2}.
These treatments assume featureless transverse superconducting modes
within superconducting nanowires and instead concentrate on the
physics of phase slips in the {\it axial} direction. Here we
complement these previous approaches by considering the effects of
{\it transverse} quantum confinement on the spatial variation of
superconducting order parameter, with consequences for the
quasi-particle excitation spectrum and the magnetic response.

\section{Formalism}

We consider a superconducting cylinder with a radius $R$ smaller than
the penetration depth $\lambda$, but much larger than the atomic scale, so
that we can describe the system with a continuum basis.  The
Bogoliubov-de Gennes (BdG) equations are
\begin{equation}
\left[ \begin{array}{cc}
  H_0 & \Delta \\
  \Delta^* & -H_0^*
\end{array} \right]
  \left[ \begin{array}{c} u \\ v \end{array} \right] =
E \left[ \begin{array}{c} u \\ v \end{array} \right]
\label{eq:bdg1}
\end{equation}
where $\Delta$ is the order parameter and $H_0$ is the Hamiltonian for
electrons,
\begin{equation}
H_0 = {1\over 2m^*}\left(-i\hbar\nabla - {e\over c}{\bf A}\right)^2 -\mu
-\mu_B\,{\bm \sigma}\cdot{\bf H}.
\label{eq:h0}
\end{equation}
Here $m^*$ is the band electron mass, $\mu$ is the chemical potential,
$\mu_B$ is the Bohr magneton, ${\bm \sigma}$ is the Pauli spin matrix
and ${\bf H}$ is the external magnetic field.  $H_0$ and its complex
conjugate $H_0^*$ act on the time-reversed electrons in Cooper
pairs. The state $[u,v]$ represents the amplitudes of the pair of
electrons which interact with each other via the pairing interaction
parametrized by the superconducting order parameter $\Delta$.  For an
axial magnetic field, ${\bf H}=H\hat{\bf z}$, the vector potential in
the Coulomb gauge is:
\begin{equation}
{\bf A}={1\over 2}rH\hat{\bf \theta},
\end{equation}
where $\hat{\bf \theta}$ is an unit vector along the azimuthal
direction.  We assume that the radius of the nanorod is sufficiently
below the penetration depth $\lambda$ that screening of the magnetic
field due to demagnetization is negligible.  In a cylindrical
coordinate system $H_0$ becomes~\cite{maki}:
\begin{eqnarray}
H_0 & = 
    & -{\hbar^2\over 2m^*}\left( {\partial^2\over\partial r^2}+{1\over r}
      {\partial\over\partial r}+{1\over r^2}{\partial^2\over\partial \theta^2}
      +{\partial^2\over\partial z^2}\right) \nonumber \\
    & &
      +\,{1\over 2}i\hbar\omega_c{\partial\over\partial \theta}
      +{1\over 8}m^* \omega_c^2 r^2-\mu -\mu_B\,\sigma_z H\\
    & = & K^0+{1\over 2}(i{\partial\over\partial \theta}-\sigma_z)
        \hbar\omega_c+{1\over 8}m^* \omega_c^2 r^2,
    \label{eq:h0r}
\end{eqnarray}
where $K^0$ is the kinetic energy in zero external field (absorbing
the chemical potential) and $\omega_c=eH/m^*c$ is the cyclotron frequency.  
With the typical separation of variables, the electron pair for
the basis in Eq.~(\ref{eq:bdg1}) consists of time-reversed electrons
in states $(mjk\uparrow)$ and $(-mj-k\downarrow)$ with $m$ the azimuthal
quantum number in $e^{im\theta}$ and $k$ the $z$-wavevector in $e^{ikz}$.  
We explicitly
write down the BdG equations by expanding $u({\bf r})$ and $v({\bf
r})$ in terms of the eigenfunctions of $K^0$ as
\begin{eqnarray}
  u^k_{mj}({\bf r}) & = &
                   u^k_{mj}\phi_{mj}(r){e^{im\theta}\over\sqrt{2\pi}}
                   {e^{ik z}\over\sqrt{L}} \\
  v^{k}_{mj}({\bf r}) & = &
                   v^k_{mj}\phi_{mj}(r){e^{im\theta}\over\sqrt{2\pi}}
                   {e^{ik z}\over\sqrt{L}}, 
\end{eqnarray}
with $L$ the length of the cylinder.  We apply a boundary condition
\begin{equation}
u({\bf r})=v({\bf r})=0 \mbox{ with }|{\bf r}|=R,
\label{eq:bc}
\end{equation}
so that the wavefunction vanishes outside the cylinder. 
This relation only imposes the condition that there are no electrons
outside the cylinder and does not make any assumptions on the
coarse-grained superconducting order parameters as usually treated in
Ginzburg-Landau theory~\cite{degennes}.
The radial term
\begin{equation}
\phi_{mj}(r)=\frac{\sqrt{2}}{RJ_{m+1}(\alpha_{mj})}J_m\left(
             \frac{\alpha_{mj}r}{R}\right),
\end{equation}
where $J_m$ is the $m$-th order Bessel function and $\alpha_{mj}$ is
its $j$-th zero.  The operator $K^0$ is diagonal with matrix elements
$K^0_{mjk}={\hbar^2/2m^*}\left(\alpha_{mj}^2/R^2+k^2\right)$.  The BdG
equation requires evaluation of matrix elements for $\langle r^2
\rangle$ and $\Delta({\bf r})$.  We will consider only the case of
order parameters with zero net angular momentum and zero net momentum
along $z$-axis, namely $m+m'=0$ and $k+k'=0$ in the product 
$u^k_{mj}v^{k'}_{m'j'}$ for order parameter $\Delta({\bf r})$. 
A paired state with finite net (angular) momentum has
higher kinetic energy than the stationary solution, and is therefore
disfavored~\cite{finiteL}.  With this choice of order parameter, we can
compute the matrix elements for $\Delta({\bf r})$ and $r^2$ as
\begin{eqnarray}
\Delta_{m;jj'} & = & \int_0^R \phi_{mj}(r)\Delta(r)\phi_{mj'}(r)rdr,\\
r^2_{m;jj'} & = & \int_0^R \phi_{mj}(r)r^2\phi_{mj'}(r)rdr.
\end{eqnarray}
The transverse modes (indexed with $mj$) are decoupled from the
longitudinal modes (indexed with $k$) and the gap equation is
simplified.

The BdG equations Eq.~(\ref{eq:bdg1}) now become (with
$I_{m;jj'}\equiv {1\over 2}m^* r^2_{m;jj'}$):
\begin{eqnarray}
& & 
\left[K^0_{mjk}-{1\over 2}(m+1)\hbar\omega_c\right]u^{k}_{mj}
\nonumber \\
& + & \sum_{j'}\left[{1\over 4}\omega_c^2 I_{m;jj'}u^{k}_{mj'}
+\Delta_{m;jj'}v^{k}_{mj'}\right] = E^k_{mj} u^{k}_{mj} \\
& & \left[-K^0_{mjk}-{1\over 2}(m+1)\hbar\omega_c\right]v^{k}_{mj}
\nonumber \\
& + & \sum_{j'}\left[{1\over 4}\omega_c^2 I_{m;jj'}v^{k}_{mj'}
+\Delta_{m;jj'}u^{k}_{mj'}\right]=E^k_{mj} v^{k}_{mj}.
\label{eq:bdg2}
\end{eqnarray}
Note that the Zeeman terms ${1\over 2}(m+1)\hbar\omega_c$ have the
same sign for $u^{k}_{mj}$ and $v^{k}_{mj}$ since they represent the
amplitudes for time-reversed states.  The order parameter $\Delta({\bf
r})$ is self-consistently expressed by the typical gap equation
\begin{eqnarray}
\Delta({\bf r}) & = & V\sum_{mjk}[1-2f(E^k_{mj})]
            u^k_{mj}({\bf r})v^k_{mj}({\bf r})^*,
\label{eq:gap}
\end{eqnarray}
where the summation range for the eigenstates is over kinetic energies
$K^0_{mjk}$ within a window $[-\omega_D,\omega_D]$ of width twice the
Debye frequency $\omega_D$.  We use generic parameter values suitable
for conventional superconductors with $T_c$ at few
Kelvins. $\Delta$ at $T=0$ converges at large diameters as will
be shown and we use the converged value $\Delta_0=3.9$ K as the bulk limit
throughout this paper. $m^*$ is set to the free
electron mass, $\omega_D=100$ K and $\mu=10000$ K. As shown below,
The Fermi velocity is then $v_F=0.55\times 10^8$ cm/sec and the 
coherence length $\xi_0=\hbar v_F/\pi\Delta_0\sim 350$ nm in the bulk.
Results are plotted in dimensionless units in this paper.
Although we have chosen a particular set of parameters, we expect
that our conclusion will hold qualitatively for conventional
low-$T_c$ superconductors.

As the diameter $D\,(\equiv 2R)$ shrinks, the transverse kinetic
energy becomes very sensitive to the boundary condition,
Eq.~(\ref{eq:bc}).  Defining a variation of radius $\delta R$, $\delta
K^0/K^0 = -2\delta R/R$.  With a small uncertainty in radius $\delta
R=1$~\AA\, around $R=10$ nm, for example, $\delta K^0\sim 0.02
\mu\sim 200$ K, which is comparable to the Debye frequency.  We
incorporate the effects of variations in the wire diameter as a noise
in the kinetic energy:
\begin{equation}
K_{mjk}=K^0_{mjk}+s_{mjk}\left|{dK^0\over dR}\right|\delta R,
\label{eq:noise}
\end{equation}
where $s_{mjk}$ is a random number uniformly distributed in $[-1,1]$
and we set $\delta R\alt 5$\ \AA.  Due to the time reversal symmetry
of static scattering at the boundary, $s_{mjk}=s_{-mj-k}$.  Here, we
have partly taken into account the radial variations via energy
levels, leaving unchanged the basis functions and the boundary
condition.  We have sampled equi-spaced $z$-momenta in the half
Brillouine zone $[0,\pi]$ with $200-500$ points, depending on the
level of convergence required.  Note that experimental nanowire
samples to date have a significant variation in diameter along their
length, but this variation is often slow on the length-scale of the
nanowire width; the longest-wavelength variations could be subsumed
into an adiabatic treatment.

\section{Results}
The density of states of a nanorod can be expressed as the sum of
one-dimensional densities of states $N_1(E)$ displaced by transverse
energy eigenvalues $E_\alpha$,
\begin{equation}
N(E) = \sum_{\alpha}N_1(E-E_\alpha).
\label{eq:dos}
\end{equation}
Quantized transverse levels strongly affect the density of states.
Transverse modes are spaced with an average level spacing $\delta
E=1/N_2$ with $N_2$ the 2-dimensional density of states, $N_2=(\pi
R^2)(m^*/\pi\hbar^2)$.  Since $N_2$ is independent of the chemical
potential, the qualitative results do not depend on the particular
position of the chemical potential.  More details will be discussed in
subsection~\ref{sec:small}.  However, $N_1(E-E_\alpha)$ has a van Hove
singularity at $E=E_\alpha$ and therefore small changes in the
chemical potential can produce quantitatively different results.
Confinement effects become strong when the level spacing $\delta E$ is
comparable to $\Delta$, {\it i.e.,}
\begin{equation}
 R\alt \sqrt{{\hbar^2\over m^* \Delta}}.
\label{eq:confine}
\end{equation}
This condition becomes $D/\xi_0\alt 0.1$ with our parameters (and
$\Delta=3.8$ K).  Due to the singularities in $N(E)$, solutions of the
BdG equation are quite sensitive to model parameters for $D/\xi_0\alt 0.1$,
even including a moderate smearing from $\delta R$.

\subsection{Large diameters}
Superconductivity under confinement has been well studied for samples
in the limit of $\delta E\ll\Delta$ and $D\ll
\lambda$~\cite{strassler, larkin, tinkham}. For cylindrical
samples with specular boundary conditions, we can explicitly solve
Eq. (\ref{eq:gap}) in the clean limit as a function of external
magnetic field along the axis,
\begin{eqnarray}
\label{eq:gorkov}
\ln(\Delta/\Delta_0) & = & -\left(1+{1\over 2\alpha^2}\right)
          \ln\left[\alpha+\sqrt{\alpha^2-1}\right] \nonumber\\
 & & +{3\over 2}\sqrt{1-\alpha^{-2}}, \ \mbox{for}\ \alpha>1  \\
& = & 0,\ \mbox{for}\ \alpha<1\nonumber
\end{eqnarray}
where $\alpha=h/\Delta$, $h=ev_FRH/2c$, and $\Delta_0$ is the order
parameter in bulk without external field. 
Str\"assler and Wyder~\cite{strassler} have obtained a result similar 
to the above equation for spherical systems and we follow essentially the 
same derivation for cylindrical systems in Appendix~\ref{app:gorkov}.
In this solution, we
make a major assumption by ignoring the spatial variation of the 
order parameter, {\it i.e.} $\Delta({\bf r})=\Delta$.  In carrying out an
analytic calculation from Eq. (\ref{eq:gap}), we make further
approximations that the position and momentum in Eq. (\ref{eq:gap})
commute and that the terms quadratic in the field ${1\over
4}m^*\omega_c^2 I_{mjk}$ in Eq. (\ref{eq:bdg2}) are negligible. When
compared with a numerical solution of the full BdG equations, the
above approximations seem reasonable in the large $D$ regime for zero
external magnetic field, (until $D/\xi_0\alt 0.57$ for our parameter
values). 

\begin{figure}[bt]
\rotatebox{0}{\resizebox{3.0in}{!}{
\includegraphics{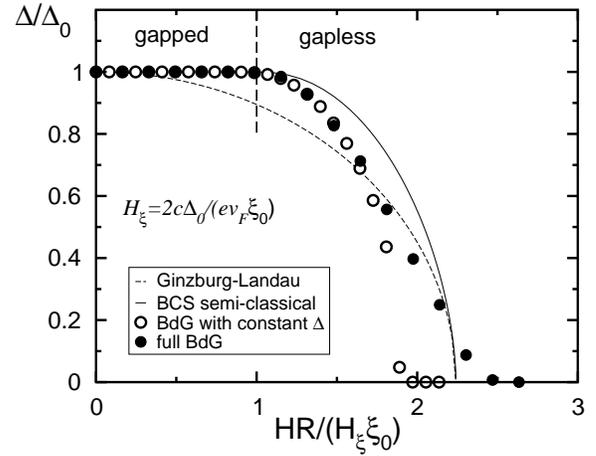}}}
\caption{\label{fig:gorkov} Averaged superconducting order parameter
in a cylindrical sample of large diameter ($D=200$ nm, $D/\xi_0=0.57$)
with axial external magnetic fields.  Normalized order parameter to
the zero field value $\Delta/\Delta_0$ is plotted as a function of
dimensionless field $HR/(H_\xi \xi_0)$ with
$H_\xi=2c\Delta_0/(ev_F\xi_0)$.  Overall agreement of the analytic
formula Eq.~(\ref{eq:gorkov}) and the numerical results are good.
Solution of Bogoliubov-de Gennes (BdG) equation in the clean limit
remains constant until $H_1=2c\Delta_0/ev_FR$.  Full solution of the
BdG equation (solid circles) has larger order parameters near the
critical field than those constrained to a spatially uniform $\Delta$
(open circle), due to spatial adjustments of superconducting
wavefunction.  }
\end{figure}

This clean-limit solution has a field dependence quite different from
the standard Ginzburg-Landau prediction. Instead of gradually decaying
from the zero-field order parameter $\Delta_0$ to zero, $\Delta$ 
stays constant up
to $H_1=2c\Delta_0/ev_FR$ and then drops to zero at the critical field
$H_c$, (see solid line in Fig.~\ref{fig:gorkov})
\begin{equation}
 H_c = {1\over 2}\exp(3/2)\,H_1= 0.454 {\Phi_0\over\xi_0 R},
\label{eq:hc}
\end{equation}
with the flux quantum $\Phi_0=hc/2e$ and the coherence length
$\xi_0=\hbar v_F/\pi \Delta_0$.
The critical field depends inversely on the diameter, since orbital
motions of electron in small samples are less influenced by magnetic field.
$H_\xi$ in Fig.~\ref{fig:gorkov} is defined as $H_1$ for $R=\xi_0$,
{\em i.e,} $H_\xi=2c\Delta_0/ev_F\xi_0$.

We have solved the BdG equations under two different conditions: first
with the constraint of a spatially uniform order parameter and then
with the constraint relaxed. The quantitative results and their
overall line shapes (in Fig.~\ref{fig:gorkov}) are in good agreement
with the analytic formula Eq.~(\ref{eq:gorkov}).  First comparing the
analytic result with the uniform-$\Delta$ calculation, we find that
the invariant order parameter up to $H=H_1$ is well reproduced. The
numerical $\Delta$ deviates downward from the analytical formula at
higher values of $H$, resulting in smaller critical field $H_c$. We
attribute the discrepancy to underestimation of the external field in
the analytic solution due to ignoring the ${\bf A}^2$ term. Further
discussion is given at the end of this subsection.

An important qualitative deviation of the full BdG solution from 
the uniform-$\Delta$ results comes at high fields near $H_c$. 
The critical field for the full BdG solution (solid circles in 
Fig.~\ref{fig:gorkov}) extends to high fields, with $\Delta$ 
decaying much slower than the usual square-root drop~\cite{sqroot}
of the uniform-$\Delta$ solution. It is clear that this phenomenon
must originate from the transverse spatial variation of the
order parameter, which we will discuss below in more detail.

\begin{figure}[bt]
\rotatebox{0}{\resizebox{3.0in}{!}{
\includegraphics{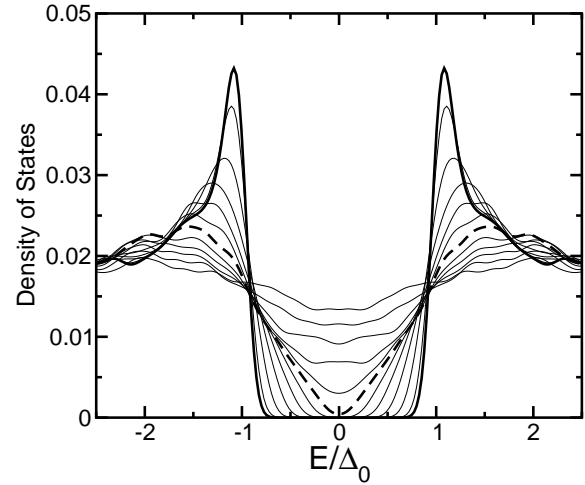}}}
\caption{\label{fig:dos} Density of states 
as a function of external field.  Superconductor becomes gapless at
$H=H_1$ (dashed line) in the clean limit well before the critical
field $H_c$, due to the coupling of orbital angular momenta to
external field.  The external fields in the plot are from $HR/(H_\xi
\xi_0)=0$ (thick line) to 1.83 with equal intervals between the
curves. }
\end{figure}

Although the average order parameter remains constant for $0<H<H_1$
(left to the dashed line in Fig.~\ref{fig:gorkov}), the density of
states reveals a closing of the excitation gap at the field $H_1$
(dashed curve in Fig.~\ref{fig:dos}), where the spectral weight at the
chemical potential becomes finite. The density of states in the gapped
region continues to increase until $H=H_c$.  In the clean limit, the
electron trajectories are not perturbed by impurity scattering and
their angular momentum ${\bf L}$ couples to the external field as
$-(e/2m^*c){\bf L}\cdot{\bf H}$. This coupling gives the term
$-{1\over 2}m\hbar\omega_c$ in Eq.~(\ref{eq:bdg2}) and contributes to
the quasi-particle excitation energy $E^k_{mj}$,
\begin{equation}
E^k_{mj} = \sqrt{{K^0_{mjk}}^2+\Delta^2}+{1\over 2}(m+1)\hbar\omega_c.
\label{eq:qpenergy}
\end{equation}
These angular momentum contributions dominate over spin contributions
and the excitation spectrum goes gapless when $\Delta\sim
(e/2m^*c)\langle {\bf L}\cdot{\bf H}\rangle \sim ev_FRH/2c$.  In
contrast, when impurity scattering dominates, the electron
trajectories are disrupted and the quasi-particle energy is no longer
in the form of Eq.~(\ref{eq:qpenergy}). In this limit, the
quasi-particle energy suffers significant level broadening and
experiences shifts only from the spin Zeeman terms~\cite{fulde}.
Therefore, away from the clean limit the density of states retains the
form of a conventional gapped superconductor~\cite{strassler}.

\begin{figure}[bt]
\rotatebox{0}{\resizebox{3.0in}{!}{
\includegraphics{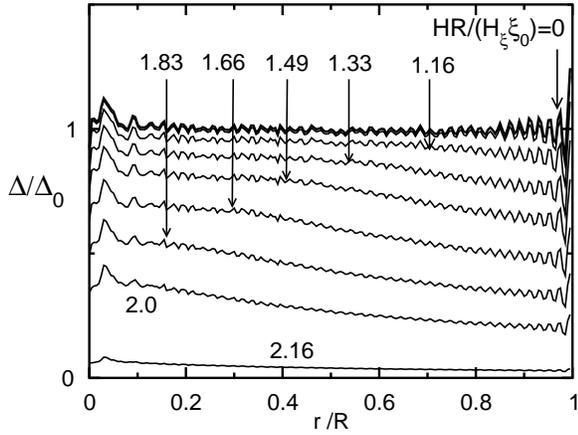}}}
\caption{\label{fig:delr} Spatial variation of order parameter $\Delta$
in the large diameter 
limit ($D/\xi_0=0.57$). For external fields of $0<H<H_1$, $\Delta$ remains flat.
As the field increases from $H_1$, $\Delta$ drops
with a knee which progresses towards ${\bf r}=0$.}
\end{figure}

At zero field, the order parameter is nearly constant over the
cylinder (see Fig.~\ref{fig:delr}), except for small oscillations and
a Gibb's phenomenon at $r=R$.  The rapid oscillations have a
wavelength proportional to $1/v_F$ and an amplitude that diminishes
for larger diameters.  Therefore we expect that these oscillations
will be averaged out on larger length scales; they are not important
for large-diameter systems. In such regime, we correctly reproduce the
boundary condition commonly used in the literature, namely a vanishing
normal derivatives of the superconducting order parameter at the
surface.  The order parameter does not change until $H=H_1$.  As $H$
exceeds $H_1$, some quasi-particle energies are pushed below zero in
Eq.~(\ref{eq:qpenergy}) and contributions from these excited
quasi-particle states reduce the order parameter in Eq.~(\ref{eq:gap})
by changing the signs of their contributions in the statistics factor,
$1-2f(E^k_{mj})$. In addition to an overall reduction of $\Delta({\bf
r})$ under external field, $\Delta({\bf r})$ also changes slope, with
a distinct knee that moves towards $r=0$ with increasing field (see
arrows in Fig.~\ref{fig:delr}).  At a radius $r$, the angular momentum
is of order $m^*rv_F$ and the order parameter begins to be suppressed
when $\Delta\sim m^*v_FrH/2c$ or $r\sim 2c\Delta/ev_FH$.  This simple
argument should be taken with caution, since the semi-classical
approximation of treating position and momentum as commutable in
Eq.~(\ref{eq:gap}) becomes worse when there is a strong spatial
variation.~\cite{spatial} At larger radius $r$, the energy difference
between the angular momenta (with $m\hbar\sim m^*r v_F$) in an
electron pair exceeds the pairing energy and therefore the pair
becomes depaired.

We emphasize that the pronounced radial dependence of order parameter
is related to the coupling of orbital angular momentum to the external
field, rather than the term ${1\over 8}m^*\omega_c^2 r^2$ in
Eq.~(\ref{eq:h0r}), which makes the superconducting order parameter
more massive. To compare these two contributions, we consider the
Ginzburg-Landau theory with an order parameter of zero total angular
momentum, as is usual in the literature~\cite{tinkham,degennes}; (see
appendix~\ref{app:GLT}).  The order parameter couples to the external
field only through ${1\over 8}m^*\omega_c^2 r^2$ and does not have any
information about the angular momenta of constituent electrons in
Cooper pairs.  The spatial dependence of $\Delta(r)$ arising from the
term $(2e/\hbar c)^2 {\bf A}^2$ turns out to be much weaker than the
angular momentum coupling in the BdG treatment.  The resulting
critical field Eq.~(\ref{app:hc}) is almost identical to the angular
momentum depairing result Eq.~(\ref{eq:hc}), as both are derived in
detail in the appendix.

It is interesting to note that, although the orbital coupling and the
${A^2}$-coupling produce very close critical fields (see
Eqs.~(\ref{eq:hc},\ref{app:hc})) in the absence of the other, they do
not act additively when both are present. When the $A^2$-term is
included on top of the orbital coupling, its effect becomes
significantly smaller than in Eq.~(\ref{app:hc}) because the orbital
effect produces fast-decaying $|\Delta(r)|^2$ near $H=H_c$.
For example, the curve labeled with $HR/(H_\xi \xi_0) =2.16$ in
Fig.~\ref{fig:delr} has $|\Delta(r=R)|^2/|\Delta(r=0)|^2\approx
1/7$. Since the $A^2\sim H^2r^2$ term couples most strongly at large
$r$, the actual $A(r)^2|\psi(r)|^2$ coupling is much smaller than the
Ginzburg-Landau theory. For instance, a reduction of the effective
$A^2$-coupling by half results in the reduction of the order parameter
by $\sqrt{1/(1+0.5)}\approx 82$\% in terms of the Ginzburg-Landau
theory when the both couplings are naively added, which nearly matches
the discrepancy in Fig.~\ref{fig:gorkov} between the
Eq.~(\ref{eq:gorkov}) (thin line) and the constant-$\Delta$ (open
circles) results.

\subsection{Small diameters}
\label{sec:small}

\begin{figure}[bt]
\rotatebox{0}{\resizebox{3.0in}{!}{
\includegraphics{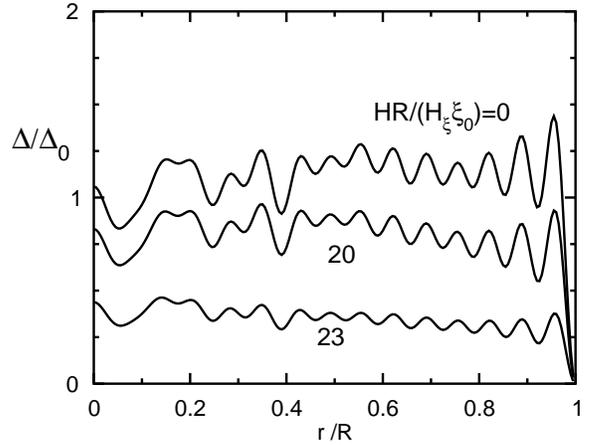}}}
\caption{\label{fig:delr20}
The order parameter $\Delta$
as a function of the radius at small diameter $D/\xi_0=0.057$
($\delta R=2$\ \AA).  The spatial variation of $\Delta$ 
is much stronger than for the large diameter wires of Fig.~\ref{fig:delr}.}
\end{figure}

When the diameter shrinks sufficiently that the transverse level
spacings $\delta E$ exceed the order parameter $\Delta$, 
the density of states on
the energy scale of $\Delta$ becomes spiky and spatial structure
arising from the transverse modes begins to show up in the radial
dependence of $\Delta({\bf r})$. Since the level spacing of the
transverse modes is inversely proportional to the effective 
mass ($\delta E=\hbar^2/m^*R^2$), 
the effect of the discrete levels will be strong for systems of small
effective mass or low transition temperature.
Fig.~\ref{fig:delr20} shows the
spatial variation of the order parameter in this regime.  A close
examination reveals two characteristic length scales in $\Delta({\bf
r})$. The shorter length scale (with rapid oscillations more apparent
at large radius) is given in terms of the Fermi velocity, i.e.,
$\Delta r\sim 2\pi\hbar/m^* v_F$. As the Fermi velocity (or the
carrier density) grows, $\Delta({\bf r})$ oscillates more rapidly.

\begin{figure}[bt]
\rotatebox{-90}{\resizebox{!}{2.0in}{
\includegraphics{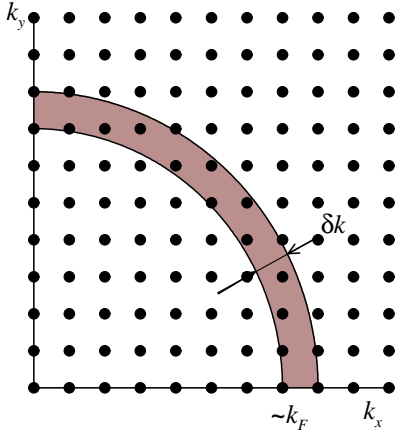}}}
\caption{\label{fig:mesh}
Schematic 2-dimensional phase space with zero axial wavevector 
($k_z=0$). 
Transverse modes within the shell of thickness $\delta k_\perp$ 
contribute strongly to the order parameter. The area of the shell 
$2\pi k_F\delta k_\perp\sim 4\pi m^*\omega_D$ does not 
depend on the choice of Fermi energy.
}\end{figure}

Apart from the structures corresponding to the Fermi wavelength, there
are more interesting and slowly varying spatial modulations,
particularly near $r=0$.  Although these modulations also appear for
large diameters (see Fig.~\ref{fig:delr}), their relative importance
grows in the smaller diameter wires.  This spatial structure arises
from the small number of transverse modes within the energy window
$[-\omega_D,\omega_D]$.  As can be seen from Eqs. (\ref{eq:gap},
\ref{app:gap}), the energy levels close to the Fermi energy contribute
strongly to the order parameter $\Delta$. Since the density of states
is peaked at the transverse energy levels (see Eq.~\ref{eq:dos}), the
resulting order parameter has larger amplitude for the states with
$\hbar^2\alpha_{mj}^2/2m^*R^2\approx \mu$, $k_z\approx 0$ and displays
the spatial characteristics of those transverse modes.  As illustrated
in Fig.~\ref{fig:mesh}, the states within the $k$-space shell of area
$2\pi k_F\delta k_\perp\sim 4\pi m^*\omega_D$ contribute most strongly
to the order parameter. Although we have used sharp energy cut-offs at $\omega_D$,
they are not expected to impose a significant quantitative change
because the weight factor in the gap equation, $u_\alpha
v_\alpha\approx \Delta/2\sqrt{\varepsilon_\alpha^2+\Delta^2}$
(cf. Eq.~(\ref{app:uv})), is small near the cutoff. The ratio of
$u_\alpha v_\alpha$ for $\varepsilon_\alpha=0$ to
$\varepsilon_\alpha=\omega_D$ is approximately $\omega_D/\Delta\approx
25$.

The shape of the order parameter is determined by which states happen
to fall into the $k$-space shell. For instance, if states of $(m=0)$
are absent in the shell, then the amplitude $\Delta({\bf r}\simeq 0)$
is depleted, since only the Bessel functions of $J_{0j}$ have
non-vanishing values at $r=0$. We caution that the order parameter is
not necessarily zero at $r=0$, since there are also states $(mjk)$
with finite $k$.  As the diameter decreases, the transverse states
become more sparse in the energy shell and the spatial structure
becomes more pronounced.  In contrast, large-diameter wires have many
contributing $(mj,k=0)$ states and the spatial variation averages out.

\begin{figure}[bt]
\rotatebox{0}{\resizebox{3.0in}{!}{
\includegraphics{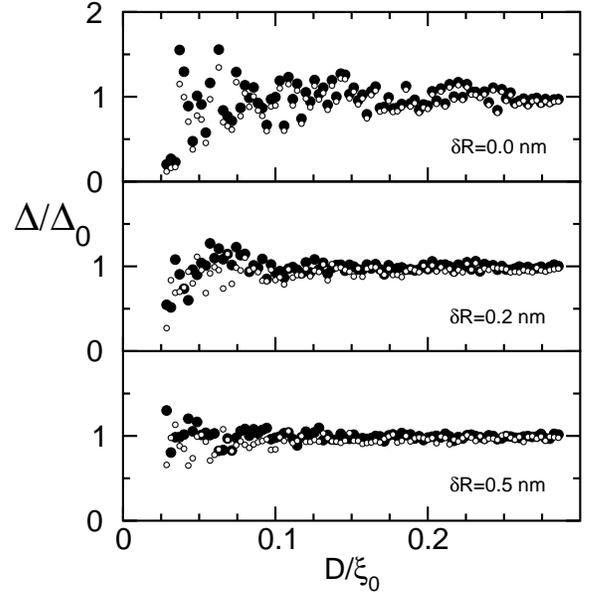}}}
\caption{\label{fig:dr}
Average order parameters $\Delta$ as a function of diameter for
different radius smearings $\delta R$.  Filled circles are the full
BdG solution while empty circles represent the constrained case where
spatial variation of $\Delta$ is disallowed.  Confinement effects
appear at $D/\xi_0\sim 0.1$, {\it i.e.,} when the transverse energy
level spacing $\delta E\approx
\Delta$.  $\Delta$ converges to about 4 K as the diameter increases, regardless
of $\delta R$.  In the full solution, $\Delta$ {\it increases}
slightly as $D/\xi_0$ drops below about 0.1. Compared to the constant-$\Delta$
behavior, this enhanced order parameter takes advantage of the spatial 
variation in the BdG solution.}
\end{figure}

Averaged order parameters $\Delta$ at zero magnetic field are plotted
as a function of diameter in Fig.~\ref{fig:dr}. The temperature is
fixed at 0.2 K and 3 different radial smearings $\delta R=0,\ 2,\ 5$
\AA\, are used.  The filled circles are solutions for full BdG
equations and the open circles impose the constraint of a constant
$\Delta$.  Regardless of $\delta R$ and the spatial constraint,
$\Delta/\Delta_0$ converges to 1 at large diameter.  $\Delta$
fluctuates considerably as the diameter decreases, with more scatter
for smaller radius smearing.  These variations arise from the sharp
van Hove singularities in the density of states at transverse
eigenvalues. Interestingly, the full BdG order parameter is
consistently larger than the constant $\Delta$ solution.  The gap
equation, Eq.~(\ref{eq:gap}) or (\ref{app:gap}), becomes particularly
simple for a constant $\Delta$ at $T=0$,
\begin{equation}
  {1\over V} = \int^{\omega_D}_0 {\rho_D(\omega)\,d\omega\over
	\sqrt{\omega^2+\Delta^2}},\ \mbox{with}\ \,
  \rho_D(\omega) = \sum_\alpha \delta(\varepsilon_\alpha-\omega),
\end{equation}
where $\varepsilon_\alpha$ is the non-interacting eigenvalue
(absorbing chemical potential).  Although the density of states
$\rho_D(\omega)$ at diameter $D$
has fluctuations due to discrete transverse energy
levels, it averages to the bulk limit.  Therefore, statistically,
$\Delta$ is expected to fluctuate about the bulk value $\Delta_0$. 
Particular lineshape of the curves (of empty circles) in
Fig.~\ref{fig:dr} is due to limited diameter sampling
and the fine-tuning of model parameters.
As the calculation of $\delta R=0.5$ nm indicates, stronger broadening 
suppresses the fluctuation of the order parameter which will converge
to $\Delta_0$ down to small $D$.
When the condition of the uniform $\Delta$ is relaxed, the $\Delta$
has the freedom to peak in regions with a higher density of electronic
states, thereby increasing the condensation energy.  For all $\delta
R$ shown here, the enhanced order parameter is most evident for
$D/\xi_0$ smaller than about 0.1, the diameter regime where the
transverse level spacing becomes comparable to $\Delta$.  For
$D/\xi_0$ of 0.05-0.1, the enhancement is roughly 10---20 \%.  
This is consistent with larger $T_c$ in small samples, as is often 
observed in thin films~\cite{strongin,dickey,goldman}.  
This tendency has been attributed
to softening of surface phonons in samples of small dimension. The
spatial variation of the order parameter $\Delta$ from confinement
effects may also contribute to this trend. 

\begin{figure}[bt]
\rotatebox{0}{\resizebox{3.0in}{!}{
\includegraphics{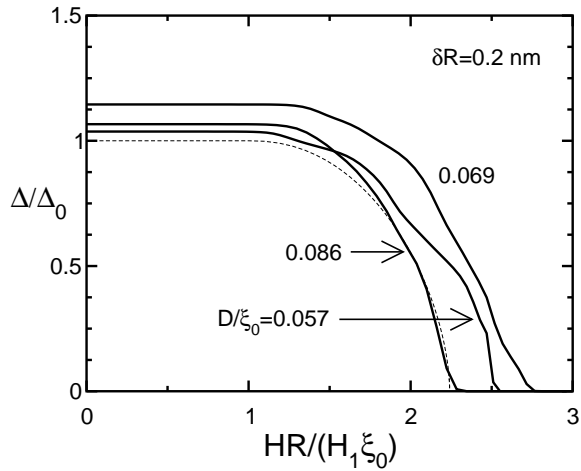}}}
\caption{\label{fig:delh}
The order parameter $\Delta$
as a function of the external field at a small diameter $D/\xi_0=
0.057-0.086$ ($D = 20 - 30$ nm, $\delta R=2$~\AA). 
$\Delta$ displays several shoulders as
the field increases and then vanishes abruptly, unlike the large
diameter case of Fig.~\ref{fig:gorkov}. This structure reflects the
discrete nature of the transverse modes. The sudden cut-off of $\Delta$
at the critical field is due to an absence of small angular momentum
transverse states near the chemical potential. The dashed line is
Eq.~(\ref{eq:gorkov}).}
\end{figure}

The order parameter $\Delta$ versus magnetic field plotted in
Fig.~\ref{fig:delh} shows shoulders that also reflect the discrete
nature of the transverse modes.  The overall shape of the curves is
similar to that for large diameters (see Fig.~\ref{fig:gorkov}).
$\Delta$ remains constant until the depairing field $e_h$ in
Eq.~(\ref{app:gap}) becomes comparable to $\Delta$.  As the field
increases further, distinctive shoulders appear. As shown in
Eq.~(\ref{app:gap}), until the condition
$\sqrt{\varepsilon_\alpha^2+\Delta^2}<e_{h\alpha}$ is satisfied for
any non-interacting state $\alpha$, the thermal factor $1-2f$ does not
change, so the gap equation yields the same $\Delta$.  At large
diameters, the number of depaired states with
$\sqrt{\varepsilon_\alpha^2+\Delta^2}>e_{h\alpha}$ increases gradually
and $\Delta$ therefore varies steadily with the external field ${\bf
H}$. However, at small diameters, the density of states has peaked
structures with van Hove singularities separated by an energy spacing
of $\hbar^2/m^*R^2$. Since only a few transverse energy peaks are
available in the energy window, the van Hove singularities have a
stronger influence on the order parameter.  Therefore kinks begin to
appear in the field dependence of $\Delta$.  As $D$ grows, these
discrete structures smooth out, as in Fig.~\ref{fig:gorkov}.

\begin{figure}[bt]
\rotatebox{0}{\resizebox{3.0in}{!}{
\includegraphics{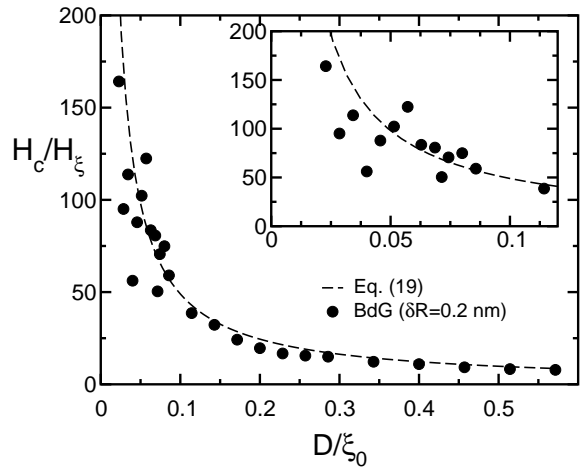}}}
\caption{\label{fig:hc_D}
The critical field vs rod diameter. The critical field $H_c$ varies
inversely proportional to the diameter $D$. The solutions of the full
BdG equation follow the general trend of Eq.~(\ref{eq:hc}), which was
obtained from the BCS equation under the constraint of a spatially
constant order parameter.  
The full BdG solution satisfies the $1/R$ prediction
down to fairly small diameters, far smaller than coherence length,
although they fluctuate strongly about the relation Eq.~(\ref{eq:hc}).}
\end{figure}

The $R$-dependence of the critical field $H_c$ is shown in
Fig.~\ref{fig:hc_D}.  Electron orbits in more tightly confined spaces
are less influenced by magnetic field, because of the smaller
depairing contribution $(e/\hbar c)\int d{\bm \ell}\cdot {\bf A}$ in
the phase of wavefunction.  The filled circles are BdG solutions with
$\delta R=2$~\AA\, and the dashed line is the semi-classical ({\it
i.e.}  $1/R$) solution of the BCS equation Eq.~(\ref{eq:hc}).  The
inset magnifies the small diameter regime. The BdG solution follows
the $1/R$ trends well down to about $D/\xi_0\sim 0.02$.  As $D$ gets
smaller, $H_c$ fluctuates substantially, but follows the overall $1/R$
behavior surprisingly well.  One of our finding is that the prediction
of the Ginzburg-Landau theory agrees remarkably well with the
microscopic solution of the BCS equation, down to small diameters
$(D\ll\xi_0)$. This conclusion may change with an inclusion of strong
interaction effects, such as increasing Coulomb interaction in strong
confinement at small diameters.

Finally, we mention that temperature dependence of the order parameter
does not show
significant deviation from the BCS results~\cite{bardeen}.  The relation
$\Delta/k_B T_c = 1.764$ holds to high accuracy for a wide range of
diameters at zero external field.

\section{conclusions}
 
We have studied the dynamics of the transverse degrees of freedom in
superconducting nanowires.  As the confinement dimension shrinks and
the level spacing becomes comparable to the order parameter $\Delta$,
the discrete nature of the transverse modes shows up in a spatial
variation of $\Delta$. In the clean limit, electronic angular momenta
are conserved and couple strongly to magnetic field to shift the
quasi-particle energy levels. This effect shows up as distinct
shoulders in the response to an external field.  In a confinement
scale comparable to or larger than the superconducting coherence
length, superconducting wavefunctions satisfy the usual boundary
condition for normal derivative $\partial\psi/\partial{\bf n}=0$ for
superconducting order parameter $\psi$. Under zero magnetic field,
$\psi$ remains constant throughout the sample, except for small and
rapid oscillations. With finite external field, $\psi$ adjust to the
vector potential with much stronger spatial variations than predicted
in Ginzburg-Landau theory.

Our results are relevant to clean-limit samples with inclusion of
level broadening effects introduced by uncertainties in diameter.  It
is useful to compare the results with the dirty-limit
theories~\cite{strassler,maki,tinkham,degennes}.  Although detailed
mechanisms for both limits are different, both systems display 
gapless behavior. The critical fields for the disappearance of order
parameters, $H_c$, behave quite differently for the two limits. In the
dirty limit~\cite{strassler}($\ell/\xi_0\ll 1$), 
$H_c$ becomes very large (before the
spin-Zeeman depairing effect dominates~\cite{fulde}), with
$H_c(\ell)\sim H_c\sqrt{\xi_0/\ell}\gg H_c$ with
$H_c$ given in Eq.~(\ref{eq:hc}) for the clean limit ($\ell/\xi_0\gg 1$).  
For the critical field $H_1$ where the excitation spectrum first becomes
gapless, $H_1/H_c|_{\ell=0}=0.954$ while
$H_1/H_c|_{\ell=\infty}=0.389$. Therefore, the clean limit remains
gapless for a wide part of the magnetic field range compared to only
4.6\% of the dirty limit. While the on-set of the order parameter
suppression and the disappearance of the excitation gap happen 
simultaneously in the clean limit, the closing of the excitation gap in 
the dirty limit happens only when the superconducting order is already
suppressed significantly.

Electrodeposition into nanoporous membranes such as polycarbonate or
anodic alumina~\cite{tian} can yield single-crystal metallic nanowires
from several different superconducting metals (tin, lead, etc.). Such
systems may be able to access the clean limit in which the phase
information of definite angular momentum states is conserved and
orbital-derived level-shifts under magnetic field become substantial.
Since the effects of discrete levels begin to appear when the level
spacing becomes comparable to the order parameter $\Delta$ (
cf. Eq.~(\ref{eq:confine})), systems of smaller $\Delta$ will exhibit
stronger confinement effects at a given wire diameter.  
Systems with small
band mass $m^*$ will have similarly strong confinement effects.

The nature of confined superconductivity in single-crystalline
metallic nanorods could perhaps be verified most clearly by the
gapless spectrum that appears at magnetic fields smaller than the
critical field. The small quasi-particle excitation energy here could
result in very interesting physics, {\it e.g.} in specific heat
measurements at temperatures below $T_c$ under external field.
Gapless superconductivity under an external field could also enhance
phase-slip rates as reflected in the electrical resistivity. Due to the
low quasi-particle excitation energies, thermal or quantum
fluctuations will overcome the condensation energy more easily.
Gapless excitations for normal electrons could contribute to
a finite residual resistance in the presence of strong scattering
at the confining surface.

\begin{acknowledgments}
We thank M. Tian, M. Chan and Y. Liu for very helpful discussions.  We
acknowledge support from the National Science Foundation DMR-0213623
and the David and Lucile Packard Foundation.
\end{acknowledgments}

\appendix

\section{Semi-classical solution for order parameter $\Delta$ in cylinder}
\label{app:gorkov}

We derive the analytic solution to the gap equation Eq.~(\ref{eq:gap})
with the constraint of a uniform order parameter $\Delta$. 
The relevant regime is when the
cylinder is large enough that the transverse level spacing is much smaller
than $\Delta$  and the external field is not too close to the critical
value.  Similar results have been derived for spherical systems using
a Green function technique.~\cite{strassler} With the assumption of a
constant $\Delta$, one approximates $[{\bf r},{\bf p}]\approx 0$,
treating ${\bf r}$ and ${\bf p}$ as independent variables. We ignore the
Zeeman term from spins, $\mu_B\bm{\sigma}\cdot {\bf H}$ in Eq.~(\ref{eq:h0}),
since it is negligible in the regime of interest. 
The term quadratic in the field, $e^2{\bf A}^2/2m^*
c^2$, is also dropped, as discussed below. 
Under these conditions, the non-interacting
Hamiltonian becomes $H_0 = \varepsilon_{\bf k}-e_h$, where $\varepsilon_{\bf
k}={1\over 2}m^* v^2-\mu$ and
\begin{equation}
e_h={e\over 2m^*c}{\bf L} \cdot{\bf H}.
\end{equation}
Solving the BdG equation Eq.~(\ref{eq:bdg1})
for a constant $\Delta$, one obtains
\begin{equation}
  E_{\bf k} = \sqrt{\varepsilon_{\bf k}^2+\Delta^2}-e_h,\ \mbox{and}\
  u_{\bf k}v_{\bf k} = {\Delta\over 2\sqrt{\varepsilon_{\bf k}^2+\Delta^2}}.
\label{app:uv}
\end{equation}
With these approximations, $u_{\bf k}v_{\bf k}$ is independent of the
field ${\bf H}$.  The gap equation can be written as
\begin{equation}
 1 = V\sum_{\bf k}{1-2f(\sqrt{\varepsilon_{\bf k}^2+\Delta^2}-e_h)\over
      2\sqrt{\varepsilon_{\bf k}^2+\Delta^2}}.
\label{app:gap}
\end{equation}
The effect of the magnetic field is reflected only in the Fermi-Dirac
function $f$.  To evaluate the angular momentum summation, we write
$e_h=(e/2m^*c){\bf L} \cdot{\bf H}=(e/2c)({\bf r}\times {\bf
v})\cdot{\bf H}=(e/2c)({\bf H}\times{\bf r})\cdot{\bf v}$.  Now, with
a fixed ${\bf r}$ and ${\bf H}$ and an isotropic distribution of ${\bf
v}$, $e_h=(ev_F/2c)rH\cos\varphi$ with the angle
$\varphi=\angle({\bf H}\times{\bf r},{\bf v})$ and $|{\bf v}|\approx
v_F$.  The summation on the right-hand side of the gap equation
Eq.~(\ref{app:gap}) becomes
\begin{equation}
\int^{\omega_D}_{-\omega_D}d\varepsilon N_0 \int_0^R{2\pi dr\,r\over \pi R^2}
    \int{d\Omega_v\over 4\pi}
    {1-2f \over 2\sqrt{\varepsilon^2+\Delta^2}},
\end{equation}
where $N_0$ is the density of states and $\Omega_v$ the solid angle
for ${\bf v}$. If $e^2{\bf A}^2/2m^* c^2$ is much smaller than the
integral limit $\omega_D$, it can be absorbed in the chemical
potential with little change in the integral $\int d\varepsilon$.  
The gap equation becomes
\begin{equation}
 1 = N_0V\int^{\omega_D}_0 d\varepsilon \int_0^1 du\,u\int_{-1}^1 d\mu
     {1-2f(\sqrt{\varepsilon^2+\Delta^2}-hu\mu)\over \sqrt{\varepsilon^2+\Delta^2}},
\label{app:int1}
\end{equation}
with $u=r/R$, $\mu=\cos\varphi$ and $h=(ev_F/2c)RH$. For $H=0$, we
have
\begin{eqnarray}
 1 & = & N_0V\int^{\omega_D}_0 d\varepsilon \int_0^1 du\,u\int_{-1}^1 d\mu
     {1-2f(\sqrt{\varepsilon^2+\Delta_0^2})\over \sqrt{\varepsilon^2+\Delta_0^2}}\nonumber \\
  & = & N_0V\ln\left({\omega_D+\sqrt{\omega_D^2+\Delta_0^2}
        \over \Delta_0}\right).
\label{app:int2}
\end{eqnarray}
Since $f=0$ for $\sqrt{\varepsilon^2+\Delta^2}>hu\mu$ in the $T=0$ limit, we
can perform the integral in Eq.~(\ref{app:int1})
\begin{eqnarray}
& & \int^{\omega_D}_0 d\varepsilon \int_0^1 du\,u\int_{-1}^1 d\mu
{f(\sqrt{\varepsilon^2+\Delta^2}-hu\mu)\over \sqrt{\varepsilon^2+\Delta^2}} \nonumber \\
 & = & {1\over 2}\left(1+{1\over 2\alpha^2}\right)\ln\left[\alpha+
       \sqrt{\alpha^2-1}\right]-{3\over 4}\sqrt{1-\alpha^{-2}}, \nonumber
\end{eqnarray}
for $\alpha=h/\Delta>1$, and 0 for $\alpha<1$.  For $\alpha>1$,
Eq.~(\ref{app:int1}) becomes
\begin{eqnarray}
\label{app:int3}
1 & = & N_0V\left\{\ln\left({\omega_D+\sqrt{\omega_D^2+\Delta^2}
        \over \Delta}\right)\right. \\
  & & \left. -\left(1+{1\over 2\alpha^2}\right)\ln\left[\alpha+
      \sqrt{\alpha^2-1}\right]+{3\over 2}\sqrt{1-\alpha^{-2}}\right\}.
      \nonumber
\end{eqnarray}
Substracting Eq.~(\ref{app:int2}) from Eq.~(\ref{app:int3}) and using
$\omega_D\gg \Delta_0$, we obtain Eq.~(\ref{eq:gorkov}).  The order
parameter $\Delta$ becomes gapless as one or both of the states in the
time-reversed pairs in Eq.~(\ref{app:uv}) are pushed out of the Debye
frequency. The external field influences $T_c$ by the statistical
factor, not directly through the field dependency in the
non-interacting density of states, as can be inferred from
Eq.~(\ref{app:int1}).  The density of states over the interaction
window of the Debye frequency remains nearly the same due to a balance
between the outflux and the influx of non-interacting energy levels
shifted by the external field.

\section{Ginzburg-Landau theory}
\label{app:GLT}

Here we consider only a stationary superconducting order parameter
($L_z=0, k_z=0$) where the Ginzburg-Landau functional for the order
parameter $\psi$ reads
\begin{eqnarray}
F & = & \!-\alpha |\psi|^2+{\beta\over 2}|\psi|^4+\left|\left(\nabla
        -i{2e\over \hbar c}{\bf A}\right)\psi\,\right|^2 \\
  & = & \!\!{H_c^2\over 8\pi}\!\!\left[-|f|^2\!+\!{|f|^4\over 2}\!+\!\xi^2\!
        \left|\!\left(\!\nabla\!-\!{im_0\omega_c r\hat{\bf \theta}\over \hbar}
        \right)\!\!f\right|^2\right]\!\!.
\end{eqnarray}
$H_c$ is the bulk critical field, $\psi=\psi_0 f\;
(|\psi_0|^2=\alpha/\beta)$, $\omega_c=eH/m_0c$, $m_0$ is free electron mass,
and $\xi$ is the
coherence length.  The Coulomb gauge is used for the vector potential
${\bf A}={1\over 2}rH\hat{\bf \theta}$.  For radius $R$ much smaller
than the penetration depth, the screening of external fields is
negligible, as assumed in the main text.  With constant $f$, we obtain
the average free energy after integration over the cylindrical
nanowire,
\begin{equation}
F={H_c^2\over 8\pi}\left[\left(-1+{1\over 2}
    \left({m_0\omega_c R\xi\over \hbar}\right)^2\right)
    f^2+{1\over 2}f^4\right].
\label{app:free}
\end{equation}
We have also numerically minimized the free energy with a spatially
varying $f$, but the spatial variation was far less significant (about
1 percent of the average order parameter for $R/\xi=0.5$) than that of
the Bogoliubov-de Gennes results discussed in the text. Therefore, the
following approximate value for the critical field is quite accurate.
Minimizing Eq.~(\ref{app:free}) over $f$ gives 
\begin{equation}
     f = \left[1-{1\over 2}\left({m_0\omega_c R\xi\over
     \hbar}\right)^2\right]^{1/2}.
\end{equation}
For vanishing $f$, the critical field is
\begin{equation}
H_c={\sqrt{2}\over \pi}{\Phi_0\over\xi R}=0.450{\Phi_0\over\xi R},
\label{app:hc}
\end{equation}
with $\Phi_0=hc/2e$.


\begin{thebibliography}{*}

\bibitem{stoner} E. C. Stoner and E. P. Wohlfarth, Philos. Trans. R. Soc.
London, Ser. A {\bf 240}, 599 (1948).

\bibitem{averin} D. V. Averin and K. K. Likharev, J. Low Temp. Phys.
{\bf 62}, 345 (1986).

\bibitem{molelec} {\it ``Molecular Electronics: Science and Technology''}, 
eds. A. Aviram and M. Ratner, Annals of the New York Academy of Sciences 
{\bf 852}, New York (1998).

\bibitem{sjohn} S. John, Phys. Rev. Lett. {\bf 58}, 2486 (1987); E. Yablonovitch, Phys. Rev. Lett. {\bf 58}, 2059 (1987).

\bibitem{graybeal} J. M. Graybeal, P. M. Mankiewich, R. C. Dynes and M. R.
Beasley, Phys. Rev. Lett. {\bf 59}, 2697 (1987).

\bibitem{ralph} D. C. Ralph, C. T. Black and M. Tinkham, Phys. Rev. Lett.
{\bf 78}, 4087 (1997).

\bibitem{tinkham} M. Tinkham, {\it Introduction to Superconductivity},
Krieger Publishing Company (1975).

\bibitem{degennes} P. G. de Gennes, {\it Superconductivity of Metals
and Alloys}, Addison-Wesley Publishing Company (1989).

\bibitem{suderow} H. Suderow, E. Bascones, A. Izquierdo, F. Guinea
and S. Vieria, Phys. Rev. B {\bf 65}, 100519 (2002).

\bibitem{misko} V. R. Misko, V. M. Fomin and J. T. Devreese,
Phys. Rev. B {\bf 64}, 014517 (2001).

\bibitem{maki_parks} K. Maki, ``Superconductivity'' Vol. II, Chapter 18,
R. D. Parks (ed.), Marcel Dekker, New York (1969).

\bibitem{strassler} S. Str\"assler and P. Wyder, Phys. Rev. {\bf 158},
319 (1967).

\bibitem{tian} M. Tian, J. Wang, J. Snyder, J. Kurtz, Y. Liu, P. Schiffer,
T. E. Mallouk and M. H. W. Chan, {\it submitted to 
Appl. Phys. Lett.}

\bibitem{giordano} N. Giordano, Phys. Rev. B {\bf 41}, 6350 (1990).

\bibitem{tinkham2} A. Bezryadin, C. N. Lau and M. Tinkham, 
Nature {\bf 404}, 971 (2000).

\bibitem{maki} M. Kato and K. Maki, J. Magn. Magn. Mater. {\bf 226},
280 (2001).

\bibitem{finiteL} Inclusion of finite angular momentum 
($L_z=\pm \hbar$) components in the BdG equation produced far smaller 
amplitudes than the zero angular momentum component. See also
Ref.~\cite{misko}.

\bibitem{larkin} L. Larkin, Zh. Eksperim. i Teor. Fiz. {\bf 48}, 232
(1965) [English transl.: Soviet Phys.-JETP {\bf 21}, 153 (1965)].

\bibitem{sqroot} Eq.~(\ref{eq:hc}) gives a steeper drop of $\Delta$ 
as $h_c-h\approx -(\Delta^2/2h_c)\ln(h_c/\Delta)$ than typical
$\Delta/\Delta_0=\sqrt{1-h^2/h_c^2}$ ($h_c=2.24\Delta_0$).

\bibitem{fulde} P. Fulde, Adv. Phys. {\bf 22}, 667 (1973).

\bibitem{spatial} If the commutation relation between position and momentum
is ignored,~\cite{larkin,strassler} we always have 
$\Delta({\bf r}=0)=\Delta_0$ at arbitrary field since the 
vector potential is zero.

\bibitem{tail} For the curve of $HR/(H_\xi\xi_0)=2.16$ in Fig.~\ref{fig:delr},
$|\Delta(r)|^2$ decays nearly by a factor of 7 at $r=R$.

\bibitem{strongin} M. Strongin, O. F. Kammerer, J. E. Crow, R. D. Parks,
D. H. Douglass Jr. and M. A. Jensen, Phys. Rev. Lett. {\bf 21},
1320 (1968).

\bibitem{dickey} J. M. Dickey and A. Paskin, Phys. Rev. Lett. {\bf 21}
1441 (1968).

\bibitem{goldman} H. M. Jaeger, D. B. Haviland, B. G. Orr and A. M. Goldman,
Phys. Rev. B {\bf 40}, 182 (1989).

\bibitem{bardeen} J. Bardeen, Rev. Mod. Phys. {\bf 34}, 667 (1962).

\end{thebibliography}
\end{document}